\documentclass{article}

\usepackage{arxiv}

\usepackage[utf8]{inputenc}  
\usepackage[T1]{fontenc}     
\usepackage{hyperref}        
\usepackage{url}             
\usepackage{booktabs}        
\usepackage{amsfonts}        
\usepackage{nicefrac}        
\usepackage{microtype}       
\usepackage{lipsum}		     
\usepackage{graphicx}
\usepackage[authoryear]{natbib}
\usepackage{doi}
\usepackage{wrapfig}

\usepackage{algorithm}
\usepackage{algorithmic}
\usepackage{lipsum}

\usepackage{xcolor}
\newcommand{\answerYes}[1]{\textcolor{blue}{#1}} 
 
\newcommand{\answerNA}[1]{\textcolor{gray}{#1}}

\usepackage{placeins}
\usepackage{pdfpages}

\title{Characterizing the Dynamics of Conspiracy Related German Telegram Conversations during COVID-19}

\author{\href{https://orcid.org/0009-0000-5015-7231}
    {\includegraphics[scale=0.06]{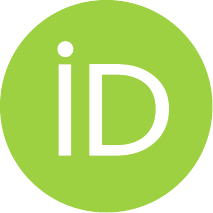}
    \hspace{1mm}Elisabeth Höldrich} \\
	IDea\_Lab\\
	University of Graz\\
	Leechgasse 34, 8010 Graz, Austria \\
	\texttt{elisabeth.hoeldrich@uni-graz.at} \\
	\And
	\href{https://orcid.org/0009-0006-4381-9261}{\includegraphics[scale=0.06]{orcid.pdf}\hspace{1mm}Mathias Angermaier} \\
	IDea\_Lab\\
	University of Graz\\
	Leechgasse 34, 8010 Graz, Austria \\
	\texttt{mathias.angermaier@uni-graz.at} \\
        \And
	\href{https://orcid.org/0000-0002-4274-4580}{\includegraphics[scale=0.06]{orcid.pdf}\hspace{1mm}Jana Lasser} \\
	IDea\_Lab\\
	University of Graz\\
	Leechgasse 34, 8010 Graz, Austria \\
	\texttt{jana.lasser@uni-graz.at} \\
        \And
	\href{https://orcid.org/0000-0003-3027-8276}{\includegraphics[scale=0.06]{orcid.pdf}\hspace{1mm}Joao Pinheiro Neto} \\
	IDea\_Lab\\
	University of Graz\\
	Leechgasse 34, 8010 Graz, Austria \\
	\texttt{joao.pinheiro-neto@uni-graz.at} \\
    }

\hypersetup{
pdftitle={Characterizing the Dynamics of Conspiracy Related German Telegram Conversations during COVID-19},
pdfsubject={German-language Telegram data and conspiracy theory research},
pdfauthor={Elisabeth Höldrich, Mathias Angermaier, Jana Lasser, Joao Pinheiro Neto},
pdfkeywords={conspiracy theories, social media, misinformation, telegram},
}

\begin{document}
\maketitle

\begin{abstract}
Conspiracy theories have long drawn public attention, but their explosive growth on platforms like Telegram during the COVID-19 pandemic raises pressing questions about their impact on societal trust, democracy, and public health. We provide a geographical, temporal and network analysis of the structure of of conspiracy-related German-language Telegram chats in a novel large-scale data set. We examine how information flows between regional user groups and influential broadcasting channels, revealing the interplay between decentralized discussions and content spread driven by a small number of key actors. 

Our findings reveal that conspiracy-related activity spikes during major COVID-19-related events, correlating with societal stressors and mirroring prior research on how crises amplify conspiratorial beliefs. By analysing the interplay between regional, national and transnational chats, we uncover how information flows from larger national or transnational discourse to localised, community-driven discussions. Furthermore, we find that the top 10\% of chats account for 94\% of all forwarded content, portraying the large influence of a few actors in disseminating information. However, these chats operate independently, with minimal interconnection between each other, primarily forwarding messages to low-traffic groups. Notably, 43\% of links shared in the data set point to untrustworthy sources as identified by NewsGuard, a proportion far exceeding their share on other platforms and in other discourse contexts, underscoring the role of conspiracy-related discussions on Telegram as vector for the spread of misinformation. 
\end{abstract}


\section{Introduction}
A conspiracy theory attempts to explain historical or ongoing events by a group of powerful people that conspires in secret to intervene in events for their own benefit and against the common good~\cite{uscinski_study_2017}.
While some scholars argue that conspiracy theories are necessary for a healthy functioning society as they play a crucial part in the marketplace of ideas where they challenge the prevailing wisdom~\cite{uscinski_study_2017}, the harmful consequences of belief in them regularly become apparent: For example, the ``great replacement'' conspiracy theory, which is grounded in racist ideology and spread by white nationalist and far-right groups, theorises that ethnically white people are purposefully being replaced by non-white people. The theory has inspired terrorists such as Anders Breivik, who murdered 77 people in Oslo and on the island Utøya in 2011~\cite{rayOsloUtoyaAttacks2011}. On the individual level, belief in ``survivalist'' conspiracy theories allegedly led to the suicide of a family in Switzerland~\cite{afpSwissPoliceBelieve2023}.
In the context of the COVID-19 pandemic, conspiracy theories related to vaccines saw increased popularity globally and became particularly salient in German-speaking countries~\cite{zehring}.

One of the most concerning aspects of conspiracy theories is their ability to act as a vector for the spread misinformation~\cite{lewandowsky_beyond_2017}.
The January 6$^\mathrm{th}$ insurrection and its close connection to belief in the QAnon conspiracy theory~\cite{bond2023rise} and the spread of misinformation around the presidential election in the U.S. highlight the threat of conspiracy theories to democratic systems and societal cohesion.
Another recent example is the COVID-19 pandemic, which has ``supercharged'' the discussion of conspiracy theories online, as particularly health-related conspiracy theories gained traction and widespread lockdowns contributed to increased online activity~\cite{Feldmann2020}. 
A recent report by the Council of Canadian Academies estimates that misinformation around COVID-19, spread in large parts by conspiracy theorists, caused 2,800 additional deaths and incurred a cost of 300 million Canadian Dollars~\cite{expert_panel_on_the_socioeconomic_impacts_fault_2023}.

Conspiracy theories have existed long before the internet~\cite{gribbin_antimasonry_1974, wood_conspiracy_1982} but, at least in public perception, they seem to have become more problematic in recent years and their spread has been tied to online communication~\cite{mahl_conspiracy_2022, theocharis_does_2021}. Therefore, one of the most pressing questions is the extent to which online communication contributes to the rapid spread of conspiracy theories, and which features of online communication allow conspiracy theories to thrive.
Here, the messenger platform Telegram has emerged as a major factor in the dissemination of conspiracy theories. The lack of moderation attracts far-right groups and communities that share conspiratorial beliefs~\cite{Bovet.2022, Curley.2022, urman_what_2022, zehring}, and use of Telegram correlates with conspiracy belief~\cite{hetzelHowCOVID19Conspiracy2022, schwaigerMindsetsConspiracyTypology2022}.
While platforms such as Telegram enable large-scale, global communication, social factors such as the need for belonging~\cite{Mashuri_Zaduqisti_2014,prooijen2017} and separation from the outgroup of the non-believers have also been identified as drivers of belief in conspiracy theories. These mechanisms could contribute to the emergence of more localised conspiracy discussion groups, in contrast to the large-scale communication structures enabled by social media platforms.

Lastly, the affordances of online communication, allowing for rapid dissemination of information, are implicated in driving the spread of misinformation. However, on almost all major platforms, several mechanisms that contribute to information spread are entangled and hard to separate: opaque content recommendation systems, influencers, moderation practices, and potential operations by motivated outside actors all contribute to the spread of content online. Here, Telegram provides the unique opportunity to study conspiracy-related information flow and information trustworthiness without the influence of content recommendation systems and moderation.

\paragraph{Problem statement} In this context, we set out to examine conspiracy-related online discourse on Telegram during the COVID-19 pandemic in German speaking countries. To this end, we study the discourse captured in the \textit{Schwurbelarchiv}~\cite{angermaier2025schwurbelarchivgermanlanguagetelegram} -- a large corpus of predominantly German-language Telegram messages that captures a substantial part of the conspiracy-related discourse on the platform from the COVID-19 era. We focus on the structure of communication, namely the distribution of discourse between discussion groups and broadcasting channels, as well as between localised regional and national/translational chats. Furthermore, we investigate whether conspiracy-related discourse on Telegram is better characterised as decentralized broader discourse among users versus discourse driven by a small number of influential actors. Lastly, we extend our analysis of communication structure by examining the relation of conspiracy discourse to the spread of untrustworthy information.  

\paragraph{Research Questions} We aim to answer the following research questions:

\begin{itemize}
    \item[\textbf{RQ1}] How has the activity in conspiracy related Telegram chats developed over time during the COVID-19 pandemic?
    \item[\textbf{RQ2}] Where are active conspiracy discussion chats located and how is the discourse distributed between regional and national/transnational chats?
    \item[\textbf{RQ3}] How is conspiracy-related discourse on Telegram better characterised: as a decentralised, collective engagement among users or as a discourse predominantly shaped by a few influential actors?
    \item[\textbf{RQ4}] How prevalent are links to untrustworthy news sources in the conspiracy-related discourse on Telegram? 
\end{itemize} 

\section{Related Works}
Telegram has become a prominent platform for the dissemination of misinformation and conspiracy theories, largely due to its lack of content moderation. The absence of systematic oversight of the content being shared on the platform enables the unhindered proliferation of unverified content. As a result, Telegram serves as a hub for conspiracy theorists, far-right actors, and counterpublics, who use its features to build information networks. Understanding the dynamics of conspiracy theories on Telegram is therefore critical for addressing questions about their spread, localisation, and interconnectivity.

Quantitative studies investigating conspiracy theories on Telegram are rare~\cite{mahl_conspiracy_2022}. The studies that do exist can be broadly divided into two categories: studies that purely focus on an analysis of the network structure between Telegram chats by inspecting forwarded messages~\cite{Bovet.2022, peetersTelegramDigitalMethods2022, urman_what_2022,willaertDisinformationNetworksQualiquantitative2022}, and studies that focus their analysis on the content of the discussions~\cite{Curley.2022,al-rawiDelegitimizingLegitimateDark2022,bodrunovaDynamicsDistrustAggression2022,hoseini_globalization_2021,gerardgillFascistCrosspollinationAustralian2021,lamorgiaUncoveringDarkSide2021,zehring,schletteOnlineStructureDevelopment2022,schulzeFarrightConspiracyGroups2022,verganiHateSpeechTelegram2022,weigandConspiracyNarrativesProtest2022} -- predominantly via employing some form of topic modelling. Many of these studies focus on the analysis of a specific and small selection of conspiracy theories or topics such as anti-vaccination sentiment~\cite{bodrunovaDynamicsDistrustAggression2022,Curley.2022,gerardgillFascistCrosspollinationAustralian2021,schletteOnlineStructureDevelopment2022,weigandConspiracyNarrativesProtest2022} and QAnon~\cite{schulze_far-right_2022, hoseini_globalization_2021}. These studies analyse content from a single~\cite{verganiHateSpeechTelegram2022} or very few ($<50$)~\cite{weigandConspiracyNarrativesProtest2022, schulzeFarrightConspiracyGroups2022, schletteOnlineStructureDevelopment2022,gerardgillFascistCrosspollinationAustralian2021,Curley.2022,bodrunovaDynamicsDistrustAggression2022,al-rawiDelegitimizingLegitimateDark2022} Telegram chats. Interestingly, some countries such as the Netherlands~\cite{willaertDisinformationNetworksQualiquantitative2022,schletteOnlineStructureDevelopment2022,peetersTelegramDigitalMethods2022} and Germany~\cite{gunzAntisemiticNarrativesYouTube2022,schulzeFarrightConspiracyGroups2022,weigandConspiracyNarrativesProtest2022,zehring} are featured significantly more often in the existent literature than others such as Italy~\cite{verganiHateSpeechTelegram2022}, Russia~\cite{bodrunovaDynamicsDistrustAggression2022}, Ireland~\cite{Curley.2022}, Australia~\cite{gerardgillFascistCrosspollinationAustralian2021}, Canada~\cite{al-rawiDelegitimizingLegitimateDark2022}, and the United Kingdom~\cite{Bovet.2022}. A small number of large-scale studies stand out, such as the~\citet{hoseini_globalization_2021} study on the spread of the QAnon conspiracy theory in a multilingual data set of 161 chats and the gigantic effort by~\citet{lamorgiaUncoveringDarkSide2021} to characterise ``dark content'' such as fakes, clones, scams, and conspiracy movements on Telegram in general in 35,382 chats.

Closest to the work presented here is the study by~\citet{zehring} which analysed the communication networks of the German Querdenken movement, the leading mobilisers of anti-COVID protests in Germany. The study identified a total of 578 conspiracy-related public Telegram chats, resulting in a dataset of of $6,294,955$ messages spanning the time from October 28, 2015, to January 3, 2022. Their analysis revealed close ties between Querdenken and far-right actors, as well as the dissemination of content related to QAnon and COVID-19 conspiracy theories. With the use of network analysis and structural topic modelling, the study highlighted the role of Telegram as a key infrastructure for radicalising communication, fostering mistrust, and amplifying far-right narratives. 

A similar, more extensive (unpublished) dataset was collected by~\citet{mohrthesis}, focusing on the diffusion of COVID-19-related (mis-)information. Recognising that snowball sampling alone can generate an overwhelming amount of chats, they used a guidance system to prioritise chats which provided relevant content, based on previously collected chats. Using a keyword list translated into 44 languages, they specifically targeted chats sharing misinformation about COVID-19. This dataset comprises of over 433,000 chats, with complete data from 128,000 chats containing over 2 billion messages posted  between January 2020 and January 2023, mainly in Russian, English and German. The content of the messages largely reflects activity from European time zones and key events such as the European COVID-19 vaccination campaign and the Russian invasion of Ukraine. The 23,000 predominantly German chats from this dataset were made available to the authors of the present study for comparative analysis with the \textit{Schwurbelarchiv}. 

There is also research exploring related but distinct aspects of Telegram such as the economic motivations in conspiracy-related chats. \citet{imperati2024conspiracymoneymachineuncovering} investigated the ecosystem of conspiracy-related Telegram chats, identifying over 17,000 such channels. Their findings highlight the monetisation strategies used by these chats, such as e-commerce, affiliate links, and crowdfunding campaigns, which collectively generate millions of dollars. This study underscores the economic motivations behind conspiracy dissemination and the use of Telegram’s environment to maintain and amplify these networks.

The above-mentioned studies show the variety of studied conspiracy theory ecosystems on Telegram, ranging from information networks and flow to their financial incentives.
However, none of the studies attempts to quantify the overall prevalence of conspiracy-related discourse on Telegram in a given temporal and regional context, or to differentiate between national/transnational discourse versus finer-grained regional activities that are rooted in local contexts. In addition, no study attempts to quantify the quality of the information shared in these discussions.

\section{Methodology}
\subsection{Collection and Processing of Data}
The \textit{Schwurbelarchiv} is a comprehensive dataset, in its raw form containing around 24\,TB of information scraped from various public German-speaking Telegram chats, covering the time span from September 23, 2015, to August 5, 2022. The dataset was originally deposited in the Internet Archive\footnote{\url{https://archive.org/details/schwurbel-archiv}}\footnote{ \url{https://schwurbelarchiv.wordpress.com/}} by an anonymous individual, and then further cleaned and processed\footnote{Cleaning and preprocessing was performed by the authors of the present work.} The data set has been deposited at [URL deleted for anonymity]. The data was gathered using a Windows Virtual Machine, with four separate Telegram Desktop sessions. Chats were then exported using Remote Desktop Protocol. Chats included in the data set were selected following a snowball sampling approach with a human-in-the-loop component to limit the number of included chats, focusing on conspiracy-related content. We obtained IRB clearance for research with and re-publication of the data contained in the \textit{Schwurbelarchiv} (vote of the IRB of Graz University of Technology from March 28, 2023).


Before analysing a dataset and conducting research, it is crucial to validate its completeness and representativeness with respect to the temporal and geographical scope studied. In~\citet{angermaier2025schwurbelarchivgermanlanguagetelegram} we provide an assessment of the completeness of the data set based on the coverage of chats in comparison to other similar data sets and the analysis of forwarded messages. We show that the \textit{Schwurbelarchiv} likely contains about 50\% of the relevant discourse on Telegram for German-language Telegram chats during the COVID-19 pandemic. An assessment of the completeness regarding geographical scope is provided in Section ``Geographical Communication Structures'' below.

\subsection{Identification of Groups and Channels}


By Telegram's platform design, Telegram chats can have two formats. Broadcast channels, which are used primarily for one-way, broadcast-style communication, feature very few actively posting authors. Typically, only the channel owner has permission to broadcast messages, although additional administrators can be added to assist with content management \footnote{Telegram Channels FAQ: \url{https://telegram.org/faq_channels?setln=uk}}. In contrast, groups facilitate many-to-many communication, where a larger number of users can actively engage in discussions. In the following, we use ``chats'' to refer to both groups and channels together, while we will use ``group'' and ``channel'' if we aim to differentiate between the two. Since we want to address the question of whether information is decentralised, emerging from broad discourse among users in groups, or if it is structurally driven by a few influential individuals in broadcast channels, it is important to be able to distinguish between channels and groups in the dataset. 

\begin{figure}[htbp]
    \centering
    \includegraphics[width=0.48\textwidth]{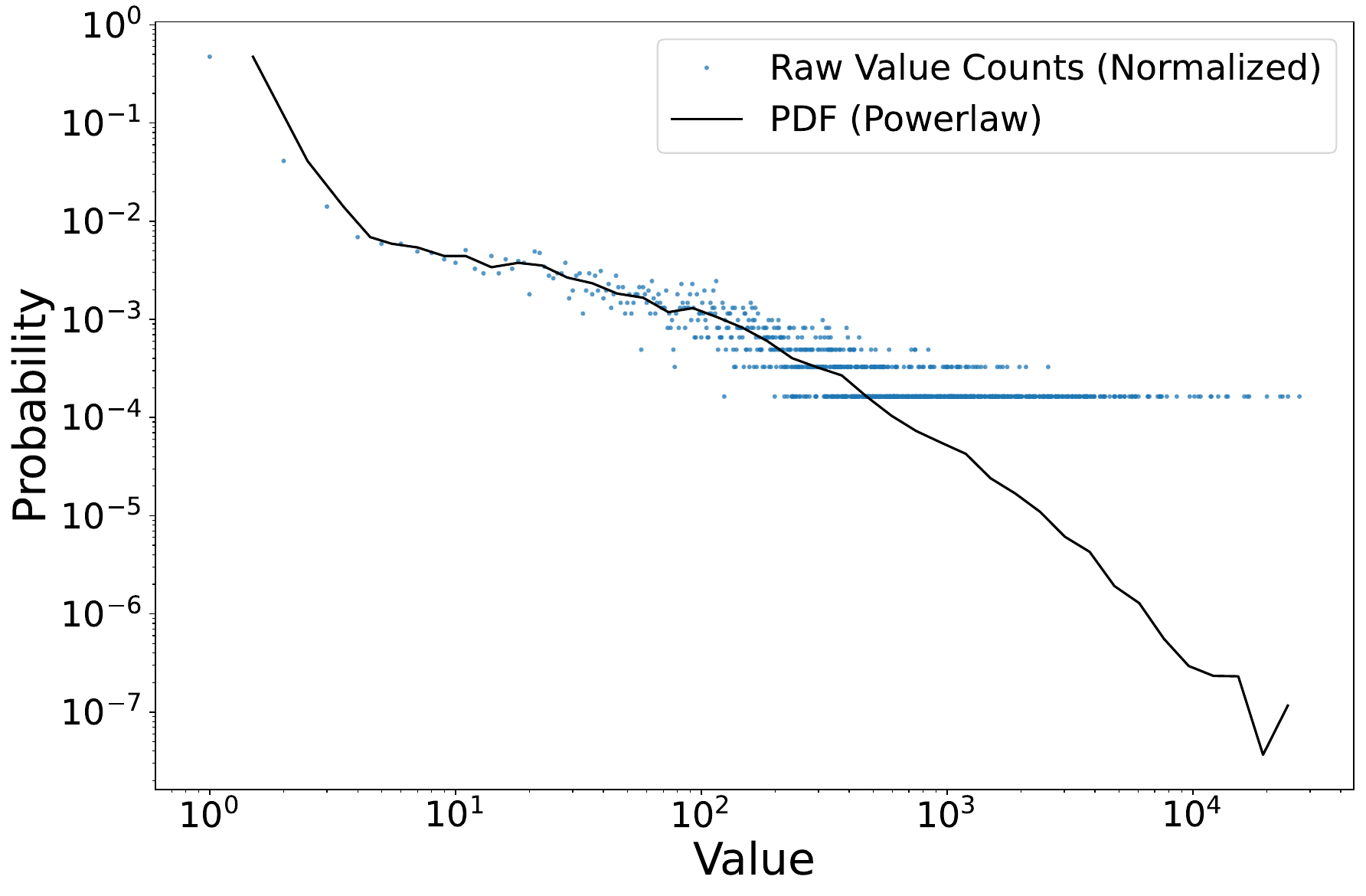}
    \caption{Log-log plot comparing the normalised raw value counts (blue) and the probability density function (PDF) of the number of authors per group in the dataset (black). }
    \label{fig:powerlaw}
\end{figure}

Since the \textit{Schwurbelarchiv} dataset does not provide information on the type of a chat, we differentiate between the two types by looking at the number of unique actively posting users in each chat. Broadcast channel administrators include the channel name in their username when posting in the channel. Thus, by examining unique usernames, we can distinguish channels from groups. If all usernames in a chat include the chat's name, we classify it as a broadcast channel. Following this approach, a slight majority of 52\% (3,130) of chats were classified as channels, while 48\% (2,946) were classified as groups. 
The distribution of the number of unique authors per chat in the dataset shown in Fig.~\ref{fig:powerlaw} reveals two distinct clusters, reflecting the underlying communication structures of groups and channels on Telegram. Groups exhibit a heavy-tailed distribution of participation, indicating many-to-many communication. 

\subsection{Identification of Geographical Scope}
Understanding the geographical scope of conspiracy theory discourse requires a systematic approach to identifying and categorising regionally and nationally focused Telegram chats. We identified regional Telegram chats within the \textit{Schwurbelarchiv} by extracting city and region names from chat titles. To this end, we used publicly available lists of towns and cities in Germany, Austria, and Switzerland. We determined matches by comparing chat names with these lists, while accounting for potential ambiguities such as abbreviations and overlapping city names (e.g., ``Traun'' and ``Traunstein''). To enhance accuracy, duplicates were excluded, though names with common substrings posed challenges. To avoid detecting substrings of city names, each name had to be preceded and followed by a space, number, punctuation, or emoji. The city and keyword lists we used are provided in the Appendix. 

For comparison, the coverage of regional chats in the \textit{Schwurbelarchiv} dataset was benchmarked against the datasets compiled by~\citet{mohrthesis} and~\citet{zehring}. The analysis included total counts of regionally identified chats, distribution across regions, and proportional representation within each dataset.
Due to the lack of spaces, punctuation, or emojis in the chat names of the other two data sets, detecting city and region names posed a challenge. The previously described approach to avoid detecting substrings or overlapping city names by forcing a space, number, punctuation, or emoji to precede and follow the city name was not possible in this case. Consequently, the estimated numbers of region-related chats may be overestimated in the datasets by~\citet{mohrthesis} and~\citet{zehring}. Despite its utility, the methodology has limitations. Abbreviations and unconventional naming conventions could lead to omissions or inaccuracies. Additionally, challenges with multilingual city names in countries like Switzerland may result in underrepresentation of chats using non-German city names. 

Next to regional chats we identified national chats by looking for chat names that contained country names. We therefore assigned chats containing either of the words ``Austria'', ``\"Osterreich'', ``Germany'', ``Deutschland'', ``Switzerland'' or ``Schweiz'' to the category of national chats. Chats that contained neither a city, region, state or country name were categorised as ``transnational''.

\subsection{Information Dissemination Networks}
Chats within the~\textit{Schwurbelarchiv} dataset are interconnected in several ways, for example by messages forwarded between them and authors posting in same chats~\cite{angermaier2025schwurbelarchivgermanlanguagetelegram}. In this study, we aim to analyse the structure of information flow through messages forwarded between chats. We therefore create a network in which each chat is a node and a directed edge between two nodes corresponds to the number of messages forwarded from one chat to the other. To create this network, we match messages that have been forwarded with original messages. To ensure identification of the correct pair of messages without a unique ID to track forwarded messages, we match \textit{messages} with \textit{fwd\_messages}, \textit{author} with \textit{fwd\_author}, and \textit{posting\_date} with \textit{fwd\_posting\_date\_message}. The underlying assumption is that one author cannot post the same message in the same second multiple times in multiple chats. The network is made available to the public here [URL deleted for anonymity] . The time difference between \textit{posting\_date} of the forwarded message and \textit{posting\_date} of the earliest occurrence of the message in the data set (e.g., the original message) further allows for analyses of information dissemination speed. 


\section{Results}
The results presented in the following sections aim to provide insights into temporal, geographical and structural characteristics of the dynamics of conspiracy-related discourse during the COVID-19 pandemic captured in the \textit{Schwurbelarchiv} dataset. 

\subsection{Temporal Dynamics}

To address our first Research Question relating to the development of the activity in conspiracy-related Telegram chats over time during the COVID-19 pandemic, we analyse the temporal patterns of message activity, number of active chats, and author participation. 

\begin{figure*}[ht]
    \centering
    \includegraphics[width=\textwidth]{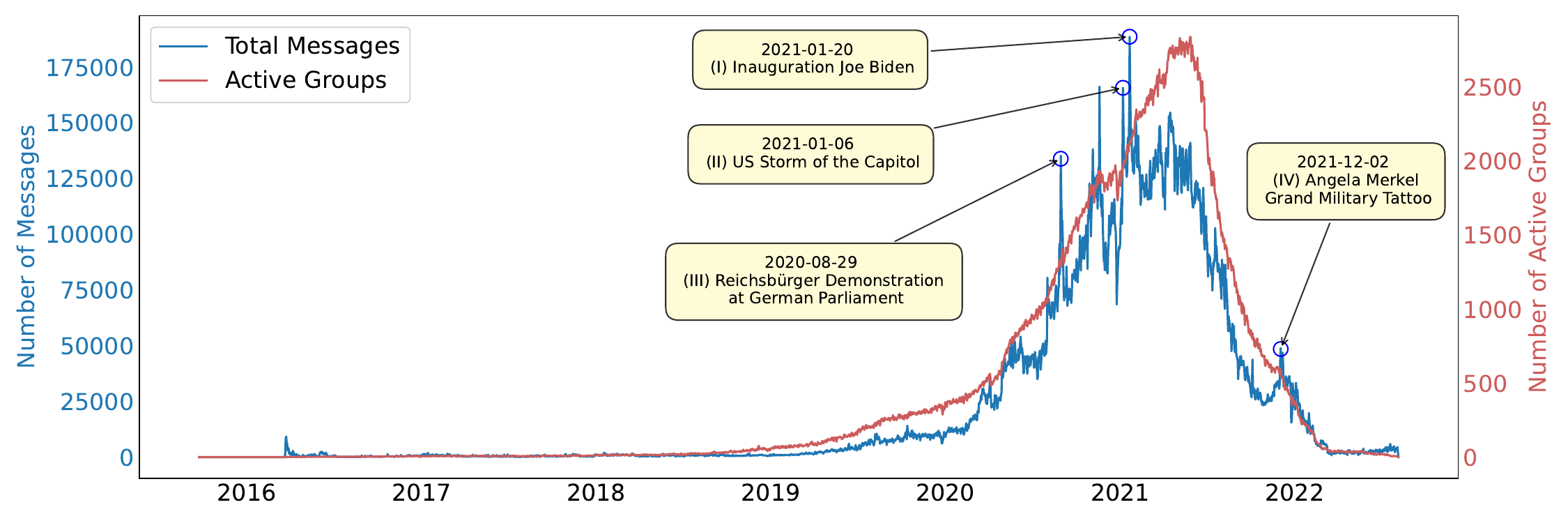}
    \caption{Number of messages (blue) and active authors (red) per day. Maxima in posting activity and related real-life events were assigned and identified manually.}
    \label{fig:active_groups}
\end{figure*}

Figure~\ref{fig:active_groups} shows the number of messages and active chats per day. Both the number of messages and the number of active chats sharply increase at the onset of the COVID-19 pandemic in Europe in early spring 2020, reaching their maximum on the observation period on January 20, 2021 -- the day of the inauguration of U.S. President Joe Biden. While there is sustained high activity throughout late 2020 and early 2021, messaging activity steeply drops toward the second half of 2021, reaching pre-pandemic levels by spring 2022. 

The timeline shows several peaks in message activity, which we matched with real-life events occurring on the same day by searching in newspaper archives \footnote{https://fazarchiv.faz.net/, https://www.welt.de/schlagzeilen/, https://www.kleinezeitung.at/suche}. On August 29, 2020, the Reichsbürger demonstration at the German parliament took place, coinciding with a significant peak in message activity. Similarly, heightened activity is visible on January 6, 2021, coinciding with the storming of the U.S. Capitol, and on January 20, 2021, during the inauguration of Joe Biden. Another significant surge occurred on December 2, 2021, corresponding to Angela Merkel's departure as German chancellor. These peaks suggest that conspiracy-related Telegram activity is strongly influenced by broader socio-political events. However, while the co-occurrence of a major event with a spike in messaging activity is indicative of a causal relationship, to definitely assign an increase in activity to a real-life event, one would have to explore the content of the messages, possibly with the use of topic modelling. Furthermore, for some of the spikes in messaging activity (e.g. the two maxima in late 2020), we were not able to identify real-life events that coincided with the increased traffic.

\begin{figure*}[ht] 
    \centering
    \includegraphics[width=\textwidth]{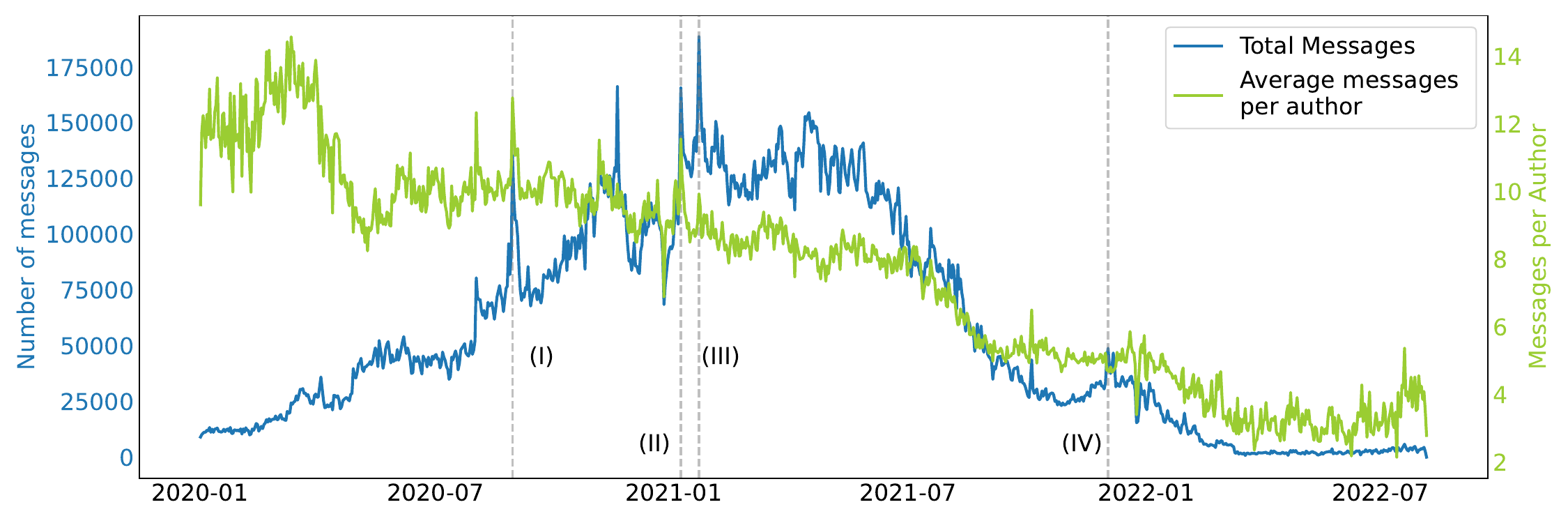} 
    \caption{Number of messages in the time period between January, 2020 and August 2022 (blue) and number of messages posted per person per day (red). Vertical dashed lines indicate the events identified in Fig.~\ref{fig:active_groups}.}
    \label{fig:authors_activity}
\end{figure*}

Next, we analyse user activity by examining the total number of messages posted per day (blue line) alongside the average number of messages posted per author (red line). Figure~\ref{fig:authors_activity} reveals a gradual decline in the average number of messages per author over time. This trend suggests a change in user engagement patterns, where individual authors post less frequently over time. This decrease may reflect several underlying factors, including a growing audience of passive participants who consume content without contributing, an increasing number of active authors spreading the total message volume more evenly, or disinterest among previously active users who reduce their participation over time. Despite the decline in individual activity, the total message volume remains substantial, particularly during the peak periods mentioned above. 

\subsection{Geographical communication structures}
The geographical scale of discourse within Telegram chats provides insights into how the dynamics of conspiracy theory dissemination are tied to regional structures of societal organisation. Analysing where conspiracy-related discussion chats are located and whether discussion chats overall are dominated by transnational or national interactions or primarily rooted in regional, small-scale communities directly addresses our second Research Question. 

In the following, we therefore first provide insights into the location of conspiracy-related chats on Telegram in Austria, Germany and Switzerland, based on our analysis of the geographical scope of chat names. In addition, we compare chats contained in the \textit{Schwurbelarchiv} to the data sets collected by~\citet{mohrthesis} and~\citet{zehring}. The comparison of datasets is summarised in Tables~\ref{tab:province_counts_austria}, \ref{tab:province_counts}, and \ref{tab:switzerland} in the Appendix, which provide detailed counts of regionally identified chats across Austrian states, German states, and Swiss cities, respectively. The aim of this comparison is to provide an assessment of the completeness of our data set on the geographical level to contextualise interpretations of geo-located messaging activity. We follow this analysis with an assessment of the flow of information between regional, national and transnational chats.

\subsubsection{Austria.}
To analyse the regional activity of conspiracy-related discourse in Austria, we assigned chats containing the name of a state, city, or town to the corresponding state for the \textit{Schwurbelarchiv} as well as the datasets studied by~\citet{mohrthesis} and~\cite{zehring}, respectively. Counts for the number of chats found for individual states and Austria as a whole are provided in Tab.~\ref{tab:province_counts_austria} in the Appendix.

In terms of individual states, Tyrol has the highest number of identified chats in the Schwurbelarchiv (13), followed by Vienna (11). The dataset by~\citet{mohrthesis} also highlights Vienna with the highest count (33), significantly surpassing other states. In contrast, the dataset by~\cite{zehring} lists only one chat each in Vienna and Austria as a whole, with no representation in other states.

The analysis also reveals notable differences in the representation of Telegram chats across the datasets. The Schwurbelarchiv dataset identifies a total of 61 chats distributed across various Austrian states, whereas the dataset by~\citet{mohrthesis} includes 197 chats, and the dataset by~\citet{zehring} contains only two. This indicates that the dataset by~\citet{mohrthesis} provides a broader coverage of Austrian regions compared to the \textit{Schwurbelarchiv}, with minimal coverage by~\citet{zehring}. In the light of these findings, differences in the number of conspiracy-related chats between individual states could as well be explained by sampling bias of chats and do not necessarily indicate a higher conspiracy-related discourse activity.  

\subsubsection{Germany.}
The vast majority of chats for which a regional or national context could be identified are related to Germany (see Table~\ref{tab:province_counts} in the Appendix for the number of chats per state and for the whole country). When comparing the number of region- or country related chats found in the \textit{Schwurbelarchiv} to the other two data sets, it is evident that the~\citet{zehring} dataset contains the fewest (30) chats related to Germany or German cities, whereas the \textit{Schwurbelarchiv} and the dataset by~\citet{mohrthesis} include a much higher number with 1,514 and 1,959 such chats, respectively. Although the \citet{mohrthesis} dataset is five times larger than the \textit{Schwurbelarchiv}, the number of regional chats included is rather similar, suggesting that the \textit{Schwurbelarchiv} proportionally samples a substantial number of regional chats. Specifically, 1,514 out of the 6,076 chats (25\%) in the Schwurbelarchiv are related to locations in Germany, in contrast to 1,959 out of 22,968 chats (9\%) in the dataset by~\citet{mohrthesis}. For~\citet{zehring}, the 30 Germany-related chats represent approximately 5\% of the data set. We therefore conclude that the representation of conspiracy-related chats on Telegram for Germany in the \textit{Schwurbelarchiv} is fairly complete.

\begin{figure*}
\centering
\includegraphics[width=\textwidth]{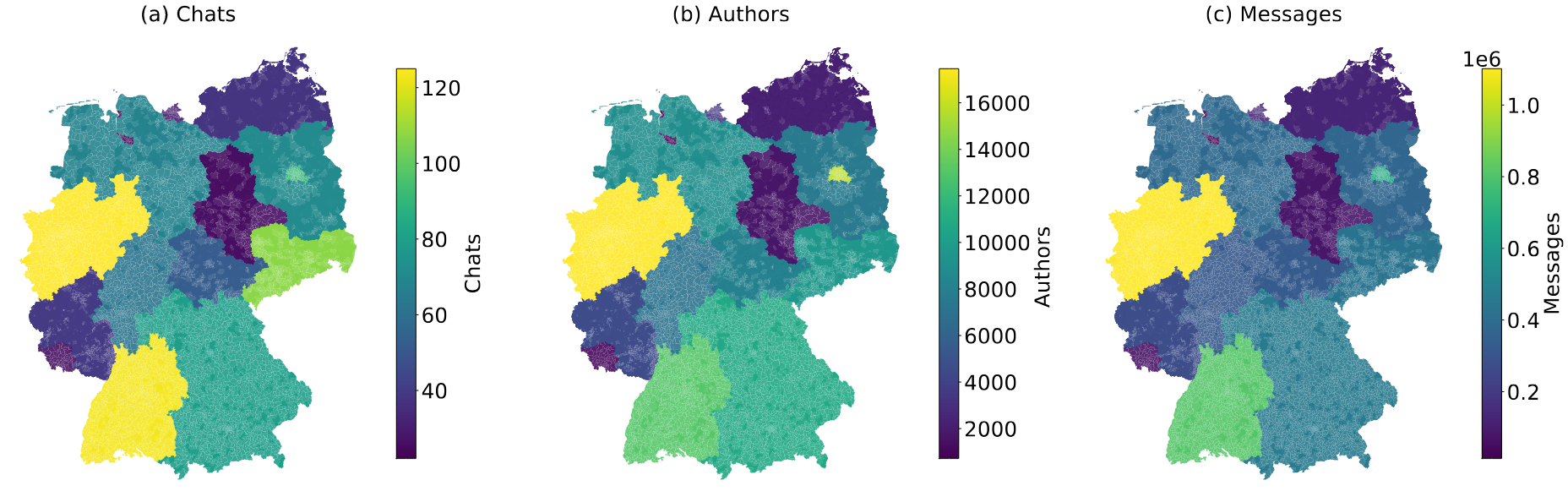}
\caption{Distribution of number of found Telegram (a) chats, (b) authors, and (c) messages in German states in the \textit{Schwurbelarchiv}.  }
\label{fig:germany}       
\vspace{-0.4cm}
\end{figure*}

Based on the distribution of chats related to each German state, Figure \ref{fig:germany} presents a map illustrating the regional distribution of chats, authors (users) and messages across Germany. The states with the highest number of chats found in the data set are North Rhine-Westphalia (Nordrhein-Westfalen) and Baden-W\"urttemberg. A similar pattern is observed in the number of authors and messages across German provinces, though these figures likely reflect the number of chats per state.

\subsubsection{Switzerland.} We attempted to conduct a similar analysis of regional chats for Switzerland. However, Switzerland poses specific challenges for such an analysis due to its multilingual nature, where cities often have different names in German, French, Italian, and Romansh. To simplify the analysis and ensure comparability across datasets, the research was restricted to the German names of cities, even though Switzerland has regions where other languages are dominant. 

For this study, we selected the ten largest cities in Switzerland and searched for chats with these names across the three datasets. Given this approach, we cannot claim any level of representativeness of the activity identified in the Telegram data for the actual conspiracy-related discourse activity in Switzerland. Therefore, the primary objective of this analysis was to determine the extent to which each dataset covers Swiss cities and to compare their relative completeness and coverage. 
Table~\ref{tab:switzerland} in the Appendix lists the cities and the number of Telegram chats associated with them in each of the three datasets. While the~\citet{zehring} dataset contains no city-specific chats and only one channel related to the country as a whole, the \textit{Schwurbelarchiv} and dataset by~\citet{mohrthesis} include 6 and 17 city-related chats respectively, while 69 and 102 chats are related to Switzerland as a whole. Even though the dataset by~\citet{mohrthesis} contains nearly five times the total number of chats, the number of chats related to Switzerland is only about twice the number of chats contained in the \textit{Schwurbelarchiv}.

\subsubsection{Regional vs. Super-regional Communication.} 
To analyse the information flow between different geographical scopes, we differentiate between three types of chats: Regional, national chats, and transnational (see Methods for details). We find 1,396 regional chats consisting of 5,611,588 messages (10\% of all messages), 254 national chats with 1,783,390 messages (3\%), and 4,426 transnational chats containing 49,729,385 messages (87\%).

To analyse the information flow, we looked at the number of messages forwarded between these different types of chats. The number of messages as well as the probability of a message being forwarded from a chat of a given type to another chat are reported in Tab.~\ref{tab:message_forwarding}. Looking at the forwarding probabilities, a few interesting patterns stand out: the most common forwarding activity is to transnational chats, which can be explained by their sheer number. While the flow of information from regional to transnational chats and vice versa is about equally likely, we see a stark difference between the probability of a message to be forwarded from a regional to a national chat (0.004\%) and from a national to a regional chat (0.2\%). This is surprising, since there are about three times as many regional chats (and messages in them) as there are national chats. This finding suggests that the discourse on the national level is more likely to inform the discourse on the regional level than the other way around. However, looking at all messages forwarded into regional chats, messages originating from national chats only make up 0.5\% of incoming messages, with the vast majority of incoming messages coming from transnational chats. From the perspective of incoming messages, regional chats are also slightly more important for national chats, where they are responsible for 1.1\% of incoming messages, with the large majority of incoming messages also originating in transnational chats.

\begin{table}[ht]
\centering
\small 
\setlength{\tabcolsep}{5pt} 
\renewcommand{\arraystretch}{1.1} 
\resizebox{\columnwidth}{!}{%
\begin{tabular}{|l|c|c|}
\hline
\textbf{Message Flow}     & \textbf{Probability (\%)} & \textbf{Messages} \\ \hline\hline
Regional → Regional       & 0.5                       & 25,358            \\ \hline
Regional → National       & 0.04                      & 2,212             \\ \hline
Regional → Transnational  & 1.1                       & 59,349            \\ \hline
National → Regional       & 0.2                       & 2,831             \\ \hline
National → Transnational  & 1.0                       & 17,782            \\ \hline
National → National       & 0.2                       & 3,274             \\ \hline
Transnational → Transnational & 7.7                   & 3,816,860         \\ \hline
Transnational → Regional  & 1.0                       & 514,992           \\ \hline
Transnational → National  & 0.4                       & 191,846           \\ \hline
\end{tabular}%
}
\caption{Probabilities and corresponding message counts for messages forwarded across different channel types.}
\label{tab:message_forwarding}
\end{table}

Furthermore, while 52\% of chats are broadcasting channels and 48\% are groups, groups generate 76\% of all messages (43,375,392 messages versus 13,744,788 messages), emphasizing the prevalence of many-to-many communication in this dataset. These findings suggest that discourse in communities plays a vital role in the discussion of conspiracy theories.

\subsection{Network Structure and Information Dissemination}
Understanding the structure of information propagation through the network of chats can offer insights into how conspiracy-related content spreads through Telegram. Here, we aim to answer our third research question of whether discourse on Telegram is predominantly shaped by a few influential actors. In this context, one focal aspect of our analysis is the assortativity of chats, based on the in- and out-degree of forwarded messages. We consider a chat strongly connected to another chat when messages are frequently forwarded from one to the other, constituting a directed link between the two chats. We interpret message forwarding as an approximation of influence between chats, representing the flow of content and ideas from one chat to the next. Here, we investigate the difference between chats with a high out-degree (or \textit{spreader} chats) and high in-degree (or \textit{receiver} chats). 

\begin{figure}[!ht]
\centering
\includegraphics[width=\columnwidth]{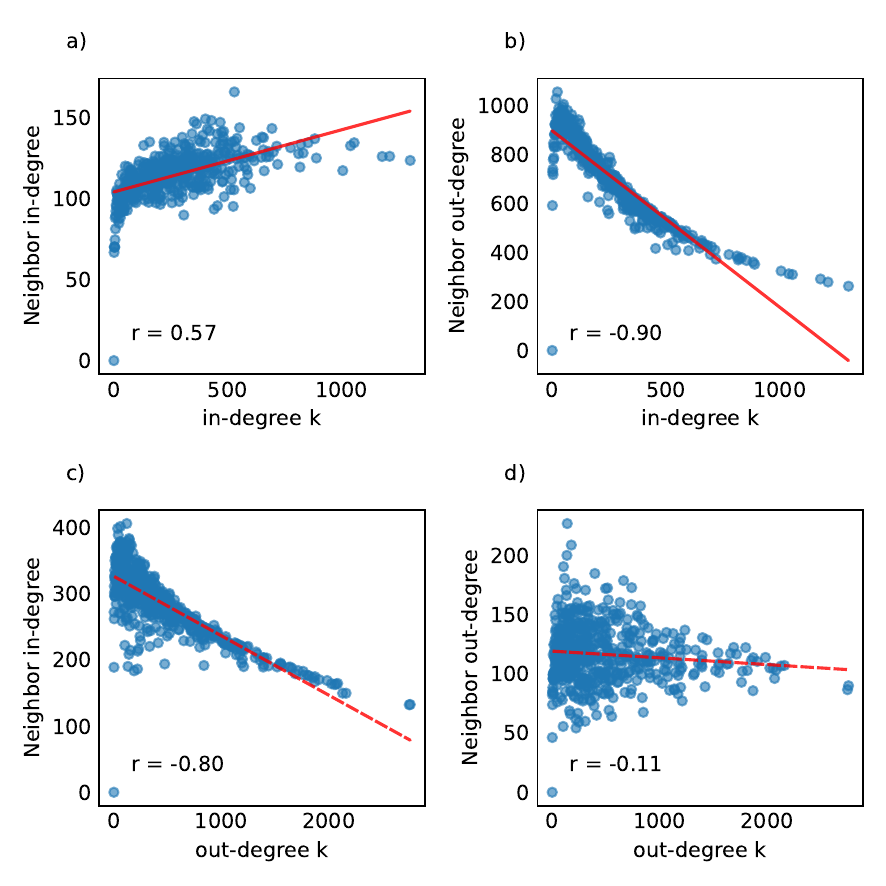}
\caption{Four combinations between the relation of a chat's in-/out-degree to the in-/out-degree of their neighbour's.}
\label{fig:assortativity}       
\vspace{-0.4cm}
\end{figure}

Figure~\ref{fig:assortativity} shows the different combinations of assortativity between a chats's in- and out-degree and the chat's neighbour's in- and out-degree. In panel a) we see a positive Pearson correlation coefficient ($\rho = 0.57$) between the in-degree of a chat and their neighbour's in-degree. This indicates that chats with a high in-degree tend to share content with other chats that also have a high in-degree and therefore \textit{receiver} chats are assortative. Panel b) shows a strong negative correlation ($\rho = -0.90$) between a chat's in-degree and its neighbours' out-degree. Consequently, chats that receive a high number of forwarded messages tend to not receive them from \textit{spreader} groups. Complementary, panel c) shows that a chat's out-degree is also strongly negatively correlated ($\rho=-0.80$) with the in-degree of its neighbours. Panel c) shows a weak anti-correlated of out-degrees ($\rho-0.11$), indicating that \textit{spreader} groups are disassortative.  Lastly, the top 10\% of \textit{spreader} chats account for 94\% of all forwarded messages.

\vspace{-0.2cm}

We interpret these findings in the following way: the \textit{Schwurbelarchiv} is a disassortative network with respect to forwarded messages. Highly influential chats with high out-degrees do not frequently share messages between each other, nor are their messages forwarded particularly often to high in-degree chats. We conclude, that messages are forwarded into chats with no predominant \textit{receiver} or \textit{spreader} characteristic. Chats do not have to be particularly strongly connected within our dataset to still be influenced by highly influential chats. Only \textit{receiver} groups are well connected with each other.

\begin{figure}
\centering
\includegraphics[width=\columnwidth]{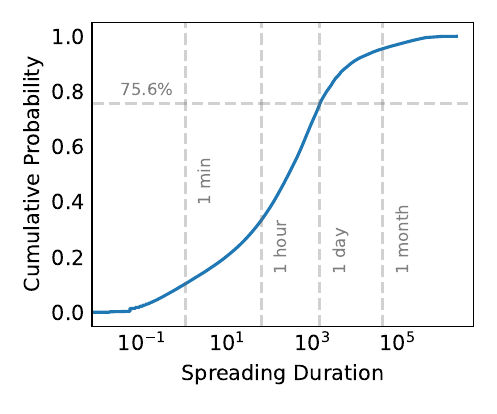}
\caption{Cumulative probability density function of forwards of a message over time. Note that forwarding time is represented on a log-scale.}
\label{fig:spread_duration}      
\vspace{-0.4cm}
\end{figure}

Next to our analysis of the influence of chats, we analyse the longevity of the influence of individual messages. Here, we use the forwarding of a message as a proxy for the influence it has and investigate for how long a message is forwarded after it was posted for the first time. Figure~\ref{fig:spread_duration} shows the cumulative probability density function of the time difference between posting the original message and forwards of that message. We find that after approximately one day, on average 75.6\% of forwards of a message will have taken place (horizontal dashed line in Fig.~\ref{fig:spread_duration}). After one month, 95.5\% of forwards do not take place anymore. A similar analysis of the number of impressions a post receives on Twitter (now ``X'') shows that after 24 hours, no relevant number of impressions can be observed for 95.5\% of all posts~\cite{Pfeffer2023}. While the forwarding behaviour analyses here is not directly comparable to the impressions  analysed by~\citet{Pfeffer2023} this indicates that the decay of influence of content on Telegram is slower than on an algorithmically moderated platform such as Twitter.

\subsection{Prevalence of Untrustworthy Information}
\begin{figure}[htbp]
    \centering
    \includegraphics[width=0.48\textwidth]{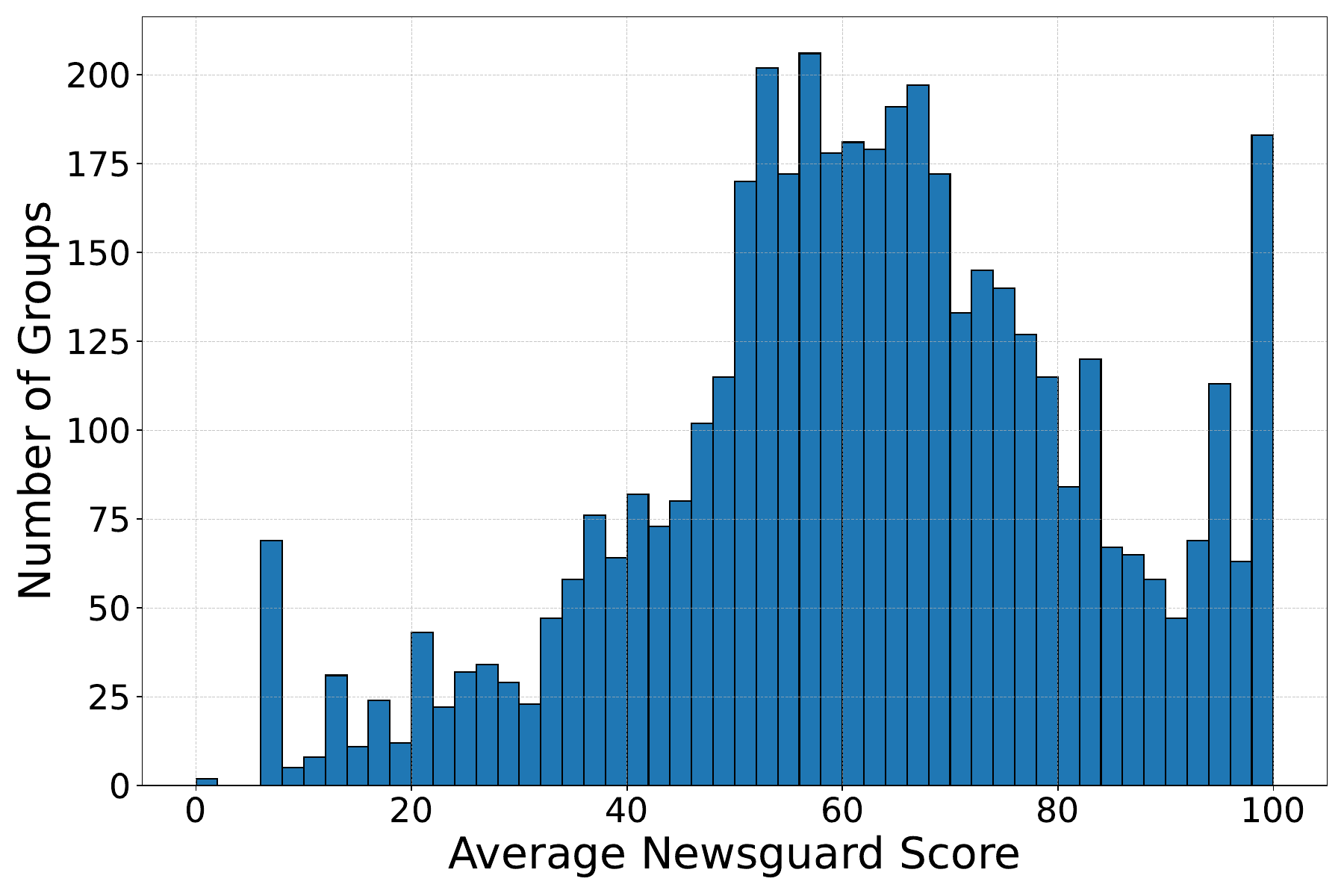}
    \caption{Distribution of average NewsGuard score per chat. Scores below 60 indicate ``untrustworthy'' domains.}
    \label{fig:newsguard_distribution}
\end{figure}

We motivated our study of conspiracy-related chats on Telegram by the tendency of conspiracy theories to act as a vector for the spread of misinformation. In this context and relating to our fourth and last research question, in this section we provide an analysis of the trustworthiness of information shared throughout the \textit{Schwurbelarchiv} data set. We operationalise misinformation as links to untrustworthy sources, using the NewsGuard data base~\cite{lin2023high, Lhring2024} to classify the trustworthiness of news outlets. Among the 2,308,880 links posted in the Telegram chats, 42.7\% point to domains classified as ``not trustworthy'' by NewsGuard (i.e., with a score below 60) \cite{NewsGuard2020}. Figure~\ref{fig:newsguard_distribution} shows the distribution of average NewsGuard scores across chats, indicating that the sharing of links to untrustworthy news sites is not a phenomenon that is isolated to a few chats, but that links to untrustworthy sites are prevalent in a majority of the chats. 

The prevalence of misinformation in online environments is currently a hotly debated topic~\cite{altay2023misinformation, Budak2024} and differences in how misinformation is classified and measured can lead to very different research outcomes~\cite{Lhring2024, Bozarth2020}. In this context, we compare the prevalence of misinformation in conspiracy related Telegram chats to the prevalence of misinformation on Twitter (now ``X''), as analysed by~\citet{lasser2022} -- a study that used the same operationalisation of misinformation. The study examined links shared by politicians in the U.S., Germany, and the United Kingdom between January 2016 and March 2022, which includes the observation period of the present study. They found that the proportion of links to domains considered ``not trustworthy'' was significantly lower. For example, Republicans in the U.S., who shared substantially more untrustworthy links than Democrats, still only posted 5.5\% of links to untrustworthy domains. Comparable figures for Germany and the United Kingdom were 8.3\% (members of the AFD) and 4.8\% (members of the DUP), respectively, for the most frequent sharers of untrustworthy links in those countries. While information sharing culture in political discourse is expected to differ from discussions of conspiracy theories, these findings underscore the very high concentration of untrustworthy information present in the Telegram chats analysed in this study.

\vspace{-0.5cm}
\section{Discussion and Conclusions}Previous studies \cite{Curley.2022,zehring} highlighted Telegram’s characteristics of minimal moderation and decentralised communication which enable the spread of misinformation. We build on these findings by analysing a new large-scale dataset, the \textit{Schwurbelarchiv}, providing a detailed view on conspiracy-related Telegram activity during the COVID-19 pandemic with a particular focus on German-speaking countries. We examine the temporal evolution of the discourse, the structural and regional patterns of the information flow, and the prevalence of untrustworthy information on the platform. As such, our study sheds light on the dynamics and characteristics of a substantial fraction of conspiracy-related discourse for a given temporal and geographical scope on Telegram. 

We show how conspiracy-related activity on Telegram increased sharply during the COVID-19 pandemic, with spikes in messaging activity correlating with societal events. This temporal pattern mirrors findings from~\citet{stein2021} who showed how crises often serve as catalysts for increased belief in conspiratorial narratives. The analysis also uncovers a substantial proportion of links to untrustworthy news sources within the dataset, as labelled so by NewsGuard. The proportion of such links on Telegram significantly exceeds figures reported in studies on other platforms such as Twitter~\citet{lasser2022}. This underscores Telegram’s role as a key hub for the spread of untrustworthy information facilitated by minimal moderation and an unregulated information-sharing environment.

Key results show that the top 10\% of chats in terms of forwarded messages account for 94\% of all forwarded content, emphasizing the significant influence of a few actors in disseminating information. Interestingly, these influential chats operate independently with minimal interconnection, primarily forwarding messages to low-traffic groups. An analysis of message longevity revealed that 75.6\% of forwards occur within 24 hours, and 95.5\% within a month. Comparatively, Twitter exhibits faster content decay, with 95.5\% of impressions occurring within 24 hours~\cite{Pfeffer2023}. This suggests that Telegram’s content influence decays more slowly, likely due to its lack of algorithmic content recommendation.

In addition to these temporal and structural dynamics, the study reveals fine-grained regional patterns in conspiracy discourse. While large-scale national and transnational discussions dominate the messaging activity, content from these chats often ends up in regional chats. This aligns with theories linking conspiracy belief to community cohesion and identity~\citet{Mashuri_Zaduqisti_2014}.

\subsubsection{Challenges, Limitations, and Directions for Future Research.} 
While the dataset used in this study is extensive, its representativeness is constrained by several factors. The data set we study contains only public chats, excluding private discussion groups and one-to-one messaging that may play a critical role in the dissemination of conspiracy theories and misinformation. Additionally, deleted messages, which are common on Telegram~\citet{buehling}, pose a significant challenge to capturing the full picture of discourse dynamics. This omission could impact the robustness of findings, particularly concerning the spread of untrustworthy information, since we would expect messages including links to untrustworthy sources to be deleted at a higher rate. Moreover, the coverage of conspiracy discussions is largely centred on German-speaking countries and the COVID-19 pandemic, limiting the generalisability of the findings to other linguistic, cultural or temporal contexts. Also, the dataset we study only covers part of the German-language conspiracy-related discourse on Telegram. As discussed in~\citet{angermaier2025schwurbelarchivgermanlanguagetelegram}, we estimate that by studying the \textit{Schwurbelarchiv}, we are able to analyse about 50\% of the relevant discourse happening on Telegram at the time. Lastly, the regional analysis provides only an estimate of localised discussions, as it relies on the matching of group names with keywords, which likely overestimates the prevalence of regional chats.

Our research and dataset opens up a number of avenues for future research, particularly concerning the content of messages, which we largely left untouched in the present study. For example, links related to cryptocurrency transfers contained in message texts allow for the analysis of monetisation dynamics within conspiracy-related Telegram chats, expanding the findings by~\cite{imperati2024conspiracymoneymachineuncovering} regarding the economic motivations of such Telegram chats. This could provide valuable insights into how financial incentives sustain and amplify conspiracy theory ecosystems.

Another opportunity for research lies in the analysis of transcribed audio content provided for the \textit{Schwurbelarchiv} (see~\citet{angermaier2025schwurbelarchivgermanlanguagetelegram} for details). Audio content on platforms like Telegram is often very information-dense, yet it remains underutilised in current research on conspiracy theories. Making use of this transcribed audio data would offer new perspectives on how conspiratorial narratives are crafted and communicated.

A comparative analysis across different platforms would also be valuable to further our understanding of how platform structure and design influence the dissemination of conspiracy-related content and misinformation. By examining differences in moderation policies, content recommendation systems, platform affordances, and user behaviours, such research could show how specific platform features contribute to the spread of these narratives. 
Such insights would also have the potential to guide public policy regarding platform governance by aiming to change the system rather than the behaviour of individual users~\cite{chaterIFrameSFrameHow2022}.

\subsubsection{Contributions and Final Thoughts.}
The prime objective of this research was to shed light on the dynamics of conspiracy-related discourse on Telegram, focusing on temporal, structural and geographical aspects as well as the trustworthiness of shared information. The motivation for our work is the need to understand societal impacts of conspiracy theories during crisis, particularly on democratic processes and public health. 
The main contributions of this research lie in its analysis of a new dataset, offering insights into the large-scale structure of conspiracy-related discourse on Telegram during the COVID-19 pandemic. The study advances the understanding of information flow between different levels of societal organisation through its analysis of the geographical scope of discussions and the relationship between content super-spreaders and discussion groups that mainly receive information. Furthermore, using NewsGuard to assess the trustworthiness of information shared, our study provides evidence for the claim that conspiracy theories act as a vector for the spread of misinformation.

\clearpage

\bibliographystyle{unsrtnat}

\bibliography{template}

\clearpage

\section{Appendix}

\begin{enumerate}

\item For most authors...
\begin{enumerate}
    \item  Would answering this research question advance science without violating social contracts, such as violating privacy norms, perpetuating unfair profiling, exacerbating the socio-economic divide, or implying disrespect to societies or cultures?
    \answerYes{Yes, this research does not violate privacy or perpetuate unfair profiling. }
  \item Do your main claims in the abstract and introduction accurately reflect the paper's contributions and scope?
    \answerYes{Yes, the abstract and introduction accurately reflect the paper’s contributions.}
   \item Do you clarify how the proposed methodological approach is appropriate for the claims made? 
    \answerYes{Yes, the methods are aligned with the claims.}
   \item Do you clarify what are possible artifacts in the data used, given population-specific distributions?
    \answerYes{Yes, we discuss limitations and potential artifacts given uncertainties around the data collection process and dataset coverage.}
  \item Did you describe the limitations of your work?
    \answerYes{Yes, limitations are discussed in the Challenges, Limitations and Directions for Future Research Section.}
  \item Did you discuss any potential negative societal impacts of your work?
    \answerYes{Yes, however for this we refer to the companion paper~\cite{angermaier2025schwurbelarchivgermanlanguagetelegram} of this study, which describes the dataset in great detail.}
    \item Did you discuss any potential misuse of your work?
    \answerYes{Yes, however for this we refer to the companion paper~\cite{angermaier2025schwurbelarchivgermanlanguagetelegram} of this study, which describes the dataset in great detail.}
    \item Did you describe steps taken to prevent or mitigate potential negative outcomes of the research, such as data and model documentation, data anonymization, responsible release, access control, and the reproducibility of findings?
    \answerYes{Yes, however for this we refer to the companion paper~\cite{angermaier2025schwurbelarchivgermanlanguagetelegram} of this study, which describes the dataset in great detail.}
  \item Have you read the ethics review guidelines and ensured that your paper conforms to them?
    \answerYes{Yes.}
\end{enumerate}

\item Additionally, if your study involves hypotheses testing...
\begin{enumerate}
  \item Did you clearly state the assumptions underlying all theoretical results?
    \answerNA{NA}
  \item Have you provided justifications for all theoretical results?
    \answerNA{NA}
  \item Did you discuss competing hypotheses or theories that might challenge or complement your theoretical results?
    \answerNA{NA}
  \item Have you considered alternative mechanisms or explanations that might account for the same outcomes observed in your study?
    \answerNA{NA}
  \item Did you address potential biases or limitations in your theoretical framework?
    \answerNA{NA}
  \item Have you related your theoretical results to the existing literature in social science?
    \answerNA{NA}
  \item Did you discuss the implications of your theoretical results for policy, practice, or further research in the social science domain?
    \answerNA{NA}
\end{enumerate}

\item Additionally, if you are including theoretical proofs...
\begin{enumerate}
  \item Did you state the full set of assumptions of all theoretical results?
    \answerNA{NA}
	\item Did you include complete proofs of all theoretical results?
    \answerNA{NA}
\end{enumerate}

\item Additionally, if you ran machine learning experiments...
\begin{enumerate}
  \item Did you include the code, data, and instructions needed to reproduce the main experimental results (either in the supplemental material or as a URL)?
   \answerNA{NA}
  \item Did you specify all the training details (e.g., data splits, hyperparameters, how they were chosen)?
    \answerNA{NA}
     \item Did you report error bars (e.g., with respect to the random seed after running experiments multiple times)?
    \answerNA{NA}
	\item Did you include the total amount of compute and the type of resources used (e.g., type of GPUs, internal cluster, or cloud provider)?
   \answerNA{NA}
     \item Do you justify how the proposed evaluation is sufficient and appropriate to the claims made? 
   \answerNA{NA}
     \item Do you discuss what is ``the cost`` of misclassification and fault (in)tolerance?
    \answerNA{NA}
  
\end{enumerate}

\item Additionally, if you are using existing assets (e.g., code, data, models) or curating/releasing new assets, \textbf{without compromising anonymity}...
\begin{enumerate}
  \item If your work uses existing assets, did you cite the creators?
    \answerYes{Yes, existing assets are cited properly.}
  \item Did you mention the license of the assets?
    \answerYes{Yes, however for this we refer to the companion paper~\cite{angermaier2025schwurbelarchivgermanlanguagetelegram} of this study, which describes the dataset in great detail.}
  \item Did you include any new assets in the supplemental material or as a URL?
    \answerYes{Yes , URL deleted for anonymity} 
  \item Did you discuss whether and how consent was obtained from people whose data you're using/curating?
    \answerYes{Yes, however for this we refer to the companion paper~\cite{angermaier2025schwurbelarchivgermanlanguagetelegram} of this study, which describes the dataset in great detail.}
  \item Did you discuss whether the data you are using/curating contains personally identifiable information or offensive content?
    \answerYes{Yes, however for this we refer to the companion paper~\cite{angermaier2025schwurbelarchivgermanlanguagetelegram} of this study, which describes the dataset in great detail.}
    \item If you are curating or releasing new datasets, did you discuss how you intend to make your datasets FAIR?
    \answerNA{NA}
    \item If you are curating or releasing new datasets, did you create a Datasheet for the Dataset?
    \answerNA{NA}
\end{enumerate}

\item Additionally, if you used crowdsourcing or conducted research with human subjects, \textbf{without compromising anonymity}...
\begin{enumerate}
  \item Did you include the full text of instructions given to participants and screenshots?
    \answerNA{NA}
  \item Did you describe any potential participant risks, with mentions of Institutional Review Board (IRB) approvals?
    \answerYes{We did not conduct experimental research with human subjects. However, we used digital trace data and obtained IRB review for our data re-publication and analysis plans (see Section Ethics and Data Protection).}
  \item Did you include the estimated hourly wage paid to participants and the total amount spent on participant compensation?
    \answerNA{NA}
   \item Did you discuss how data is stored, shared, and deidentified?
   \answerNA{NA}
\end{enumerate}

\end{enumerate}

\clearpage 

\section{City Lists Used for Regional Analysis}

The following lists include the cities and regions used to identify and categorize regional Telegram chats in the Schwurbelarchiv dataset. These lists were based on publicly available data and processed to detect regional references in group names.

\subsection{List of Austrian Cities}
The list below contains names of Austrian cities considered for regionality analysis:

\noindent
\texttt{[
'Vienna', 'Graz', 'Linz', 'Lochau', 'Salzburg', 'Innsbruck', 'Klagenfurt', 'Villach', 'Wels', \\
'Sankt Pölten', 'Krems an der Donau', 'Dornbirn', 'Wiener Neustadt', 'Steyr', 'Bregenz', 'Leonding', \\
'Klosterneuburg', 'Traun', 'Lustenau', 'Amstetten', 'Kapfenberg', 'Hallein', 'Mödling', 'Kufstein', \\
'Traiskirchen', 'Schwechat', 'Hohenems', 'Stockerau', 'Ansfelden', 'Telfs', 'Bruck an der Mur', \\
'Spittal an der Drau', 'Perchtoldsdorf', 'Bludenz', 'Eisenstadt', 'Hall in Tirol', 'Wörgl', 'Schwaz', \\
'Marchtrenk', 'Leibnitz', 'Korneuburg', 'Neunkirchen', 'Knittelfeld', 'Vöcklabruck', 'Enns', \\
'Lienz', 'Brunn am Gebirge', 'Rankweil', 'Bad Vöslau', 'Götzis', 'Gänserndorf', 'Gerasdorf bei Wien', \\
'Gleisdorf', 'Lauterach', 'Strasshof an der Nordbahn', 'Köflach', 'Laakirchen', 'Purkersdorf', \\
'Wiener Neudorf', 'Voitsberg', 'Guntramsdorf', 'Attnang-Puchheim', 'Maria Enzersdorf', 'Deutsch-Wagram', \\
'Wolfurt', 'Langenzersdorf', 'Freistadt', 'Bruck an der Leitha', 'Landeck', 'Pasching', 'Kottingbrunn', \\
'Seiersberg', 'Grödig', 'Zeltweg', 'Hafendorf', 'Jenbach', 'Vösendorf', 'Kalsdorf bei Graz', 'Völs', \\
'Hartberg', 'Altach', 'Feldkirchen bei Graz', 'Asten', 'Mattighofen', 'Gallneukirchen', 'Hörbranz', \\
'Neuhofen an der Krems', 'Hörsching', 'Gloggnitz', 'Wagna', 'Kirchbichl', 'Oberndorf bei Salzburg', \\
'Timelkam', 'Thalheim bei Wels', 'Bergheim', 'Wildon', 'Schwertberg', 'Bad Gleichenberg', 'Schärding', \\
'Lieboch', 'Leopoldsdorf', 'Stadl-Paura', 'Lenzing', 'Bad Hall', 'Sollenau', 'Grieskirchen', 'Birkfeld', \\
'Mauthausen', 'Leobersdorf', 'Bürmoos', 'Weißkirchen in Steiermark', 'Althofen', 'Bisamberg', 'Ottensheim', \\
'Oberwaltersdorf', 'Eichgraben', 'Koblach', 'Neudörfl', 'Mils bei Solbad Hall', 'Kirchdorf', 'Puchenau', \\
'Sankt Georgen im Attergau', 'Oberalm', 'Tribuswinkel', 'Felixdorf', 'Schwanenstadt', 'Hallwang', \\
'Bad Schallerbach', 'Götzens', 'Sankt Georgen an der Gusen', 'Eferding', 'Mäder', 'Gössendorf', 'Schwarzach', \\
'Gumpoldskirchen', 'Fußach', 'Pinsdorf', 'Loosdorf', 'Gratwein', 'Pfaffstätten', 'Schwarzach im Pongau', \\
'Theresienfeld', 'Ludesch', 'Lambach', 'Neufeld an der Leitha', 'Göfis', 'Fernitz', 'Hausmannstätten', \\
'Pirka', 'Kierling', 'Sankt Florian am Inn', 'Klaus', 'Brixlegg', 'Ennsdorf', 'Kematen in Tirol', \\
'Biedermannsdorf', 'Elixhausen', 'Niederndorf', 'Gallspach', 'Aldrans', 'Sankt Marein im Mürztal', \\
'Kaindorf an der Sulm', 'Rohrbach', 'Sulz', 'Raaba', 'Schlins', 'Windischgarsten', 'Bludesch', \\
'Werndorf', 'Gießhübl', 'Meiningen', 'Sistrans', 'Aschach an der Donau'
]}
\subsection{List of Austrian Regions}
The list below contains the regions and other geographical references considered for regionality analysis:

\noindent
\texttt{[
'Wien', 'Steiermark', 'Oberösterreich', 'Vorarlberg', 'Salzburg', 'Tirol', 'Kärnten', 'Niederösterreich', \\
'Burgenland', 'Kaernten', 'Oberoesterreich', 'Niederoesterreich', 'ooe', 'noe', 'oesterreich', 'oö', \\
'austria', 'vienna'
]}
\subsection{List of Swiss Cities}
The following list includes the Swiss cities considered for regionality analysis:

\noindent
\texttt{[
'Zürich', 'Genf', 'Basel', 'Lausanne', 'Bern', 'Winterthur', 'Luzern', 'St. Gallen', 'Lugano', 'Biel'
]}

\subsection{List of Swiss Regional References}
The list below contains regional and geographical references related to Switzerland:

\noindent
\texttt{[
'Schweiz', 'Switzerland', 'Schweizer', 'Swiss', 'Svizzera', 'Suisse'
]}

\subsection{Complete Combined List of Cities and Regions in Germany}
The following is the complete list of cities and regions across all German states (Bundesländer), used for regionality analysis:

\noindent
\texttt{[
'Rheinland-Pfalz', 'Alzey', 'Andernach', 'Annweiler', 'Arbach', 'Asbach', 'Astert', 'BadKreuznach', 'Berlingen', \\
'Bornheim', 'Brauweiler', 'Bretzenheim', 'Bruecken', 'Eborn', 'Ellenberg', 'Ellern', 'Enkirch', 'Ensheim', \\
'Erbach', 'Ersfeld', 'Gladbach', 'Hambach', 'Idelberg', 'Kaiserslautern', 'Koblenz', 'Landau', 'Landscheid', \\
'Langenfeld', 'Laudert', 'Lemberg', 'Mainz', 'Mendig', 'NiederOlm', 'Offenbach', 'Orbis', 'Pirmasens', 'Ramberg', \\
'Saarburg', 'Salm', 'Schoenborn', 'Schoeneberg', 'Senheim', 'Talmuehle', 'Tellig', 'Trier', 'Westheim', 'Wolfstein', \\
'Baden-Württemberg', 'Althengstett', 'Waldshut', 'WaldshutTiengen', 'VillingenSchwenningen', 'StBlasien', 'Schutterwald', \\
'Rottweil', 'Neckarsulm', 'Pfaffenhofen', 'Gerlingen', 'Nagold', 'Friedrichshafen', 'Ditzingen', 'Biberach', \\
'BadKrozingen', 'BadMergentheim', 'Balingen', 'Aichtal', 'Aidlingen', 'BadenBaden', 'Brackenheim', 'Bruchsal', \\
'Calw', 'Crailsheim', 'Dettingen', 'Eberbach', 'Forchheim', 'Freiberg', 'Freiburg', 'Freudenstadt', 'Gutach', \\
'Hechingen', 'Heidelberg', 'Heilbronn', 'Heiligenberg', 'Hohenlohe', 'Ibach', 'Illingen', 'Irndorf', 'Karlsruhe', \\
'Konstanz', 'Kornwestheim', 'Lauterbach', 'Leingarten', 'Lichtenstein', 'Loerrach', 'Ludwigsburg', 'Malsch', \\
'Mannheim', 'Marbach', 'Osterholz', 'Owen', 'Pforzheim', 'Pfullendorf', 'Pfullingen', 'Ravensburg', 'Reutlingen', \\
'Rosenberg', 'Rosengarten', 'Schramberg', 'Sigmaringen', 'Sinsheim', 'Stuttgart', 'Tengen', 'Tuebingen', 'Tuttlingen', \\
'Ulm', 'Vogt', 'Wiesloch', \\
'Nordrhein-Westfalen', 'Aachen', 'Wermelskirchen', 'Wadersloh', 'Stadtlohn', 'Remscheid', 'OerErkenschwick', \\
'Niederkassel', 'Leichlingen', 'KampLintfort', 'Gummersbach', 'Gladbeck', 'Greven', 'Gelsenkirchen', 'Erftstadt', \\
'Erkelenz', 'Erkrath', 'Emsdetten', 'BadOeynhausen', 'Detmold', 'Baesweiler', 'Bedburg', 'CastropRauxel', 'Coesfeld', \\
'Schlewig-Holstein','schleswigholstein',\\ 'Ahrensburg', 'Uetersen','Owschlag',\\'Norderstedt', 'Hohenwestedt','Harrislee',\\'Geesthacht','Barmstedt','BadOldesloe', 'BadBramstedt', 'Bargteheide', 'Buettel', 'Delve', 'Elmenhorst', 'Elmshorn', 'Flensburg', 'Hamberge', 'Hemdingen', 'HenstedtUlzburg', 'Itzehoe', 'Kiel', 'Landscheide', 'Luebeck','Oldenburg', 'Pinneberg', 'Quickborn', 'Ramstedt', 'Ratzeburg', 'Rendsburg', 'Rickert', 'Ruede', 'Schlichting', 'Talkau'
'Arnsberg', 'Beckum', 'Bielefeld', 'Bocholt', 'Bochum', 'Bonn', 'Bornheim', 'Bottrop', 'Dinslaken', 'Dorsten', \\
'Dortmund', 'Duisburg', 'Duesseldorf', 'Elsdorf', 'Enger', 'Espelkamp', 'Euskirchen', 'Geilenkirchen', 'Geldern', \\
'Guetersloh', 'Hattingen', 'Heinsberg', 'Herford', 'Herzogenrath', 'Hoexter', 'Iserlohn', 'Juelich', 'Koeln', \\
'Korschenbroich', 'Krefeld', 'Lennestadt', 'Leverkusen', 'Lippstadt', 'Lohmar', 'Luebbecke', 'Mettmann', 'Minden', \\
'Moenchengladbach', 'Monschau', 'Muenster', 'Neuss', 'Paderborn', 'Pulheim', 'Ratingen', 'Recklinghausen', 'Rheinbach', \\
'Rheinberg', 'Salzkotten', 'Velbert', 'Viersen', 'Wipperfuerth', 'Wuppertal', 'Xanten', \\
'Hessen', 'BadVilbel', 'Trendelburg', 'Pfungstadt', 'Seligenstadt', 'Frankenberg', 'Gelnhausen', 'BadHomburg', \\
'Borken', 'BadWildungen', 'Baunatal', 'Bensheim', 'Biblis', 'Biedenkopf', 'Braunfels', 'Brechen', 'Darmstadt', \\
'Dreieich', 'Erzhausen', 'Fernwald', 'Frankfurt', 'Friedberg', 'Fulda', 'Fuerth', 'Gedern', 'Griesheim', 'Hanau', \\
'Heusenstamm', 'Hoechst', 'Kassel', 'Langen', 'Lauterbach', 'Limburg', 'Marbach', 'Marburg', 'Muenster', 'Naumburg', \\
'Obertshausen', 'Offenbach', 'Rabenau', 'Rotenburg', 'Taunusstein', 'Waldeck', 'Wartenberg', 'Weilburg', 'Weimar', \\
'Wiesbaden', \\
'Mecklenburg-Vorpommern', 'Altona', 'Nevern', 'Parchim', 'Altenkirchen', 'BadDoberan', 'Bellin', 'Below', 'Born', \\
'Buelow', 'Dargun', 'Demmin', 'Godern', 'Greifswald', 'Gresse', 'Guestrow', 'Hamberge', 'Lubmin', 'Malchin', \\
'Neubrandenburg', 'Patzig', 'Quassel', 'Rostock', 'Stralsund', 'Tessin', 'Teterow', 'Usedom', 'Utecht', 'Viez', \\
'Waldeck', 'Wismar', 'Wittenburg', 'Wolgast', 'Zehna', \\
'Bayern', 'Aschaffenburg', 'Wolfratshausen', 'Schorndorf', 'Schrobenhausen', 'Memmingen', 'Miltenberg', 'Mindelheim', \\
'Monheim', 'GarmischPartenkirchen', 'Erlangen', 'Coburg', 'Deggendorf', 'Dingolfing', 'BadToelz', 'Bayreuth', \\
'Berchtesgaden', 'Aufkirchen', 'Augsburg', 'BadWoerishofen', 'Bamberg', 'Bellenberg', 'Castell', 'Chiemsee', \\
'Dillingen', 'Ergoldsbach', 'Forchheim', 'Germering', 'Guenzburg', 'Ingolstadt', 'Inning', 'Kempten', 'Kitzingen', \\
'Kraiburg', 'Landshut', 'Langenfeld', 'Lenggries', 'Miesbach', 'Muehlhausen', 'Muenchen', 'Muenster', 'Neuoetting', \\
'Neutraubling', 'NeuUlm', 'Nuernberg', 'Oberasbach', 'Passau', 'Regensburg', 'Roding', 'Schwandorf', 'Schweinfurt', \\
'Taufkirchen', 'Traunstein', 'Waldkraiburg', 'Wartenberg', 'Wunsiedel', 'Wuerzburg', \\
'Berlin', 'Bremen', 'Bremerhaven', 'Hamburg', 'HamburgInselNeuwerk', 'Neuwerk', 'Saarland', 'Beckingen', 'Bexbach', \\
'Blieskastel', 'Dillingen', 'Eppelborn', 'Friedrichsthal', 'Gersheim', 'Grossrosseln', 'Heusweiler', 'Homburg', \\
'Illingen', 'Kirkel', 'Kleinblittersdorf', 'Lebach', 'Losheim', 'Mandelbachtal', 'Marpingen', 'Merchweiler', \\
'Mettlach', 'Nalbach', 'Namborn', 'Nohfelden', 'Nonnweiler', 'Oberthal', 'Ottweiler', 'Puettlingen', 'Quierschied', \\
'RehlingenSiersburg', 'Riegelsberg', 'Saarbruecken', 'Saarlouis', 'Saarwellingen', 'Schiffweiler', 'Schmelz', \\
'Schwalbach', 'SpiesenElversberg', 'StIngbert', 'StWendel', 'Sulzbach', 'Tholey', 'Überherrn', 'Voelklingen', 'Wadern', \\
'Wadgassen', 'Wallerfangen','Weiskirchen', 'Niedersachsen', 'Barsinghausen','Northeim', 'Salzgitter','Egestorf','Buxtehude', 'Cloppenburg' 'Bodensee', 'Braunschweig','Cappel', 'Celle', 'Cuxhaven', 'Dahlenburg', 'Delmenhorst', 'Diepholz', 'Freiburg', 'Garbsen', 'Gifhorn', 'Goettingen', 'Hannover','Heidenau','Hildesheim', 'Lueneburg', 'Nienburg', 'Oldenburg', 'Osnabrueck', 'Rotenburg' ,'Seevetal', 'Thedinghausen', 'Uelzen', 'Uetze', 'Verden','Wildeshausen', 'Wremen', 'Sachsen','BadDueben','Oberlungwitz', 'Frankenberg','Colditz','BrandErbisdorf', 'BadLausick', 'Bautzen', 'Chemnitz', 'Coswig', 'Dahlen', 'Delitzsch', 'Dippoldiswalde', 'Doebeln', 'Dresden', 'Goerlitz', 'Groitzsch', 'Hainichen', 'Heidenau', 'Kalbitz', 'Leipzig', 'Leisnig', 'Lunzenau', 'Oschatz', 'Possendorf','Satzung','Zeughaus', 'Zwenkau', 'Zwickau', 'Zwoenitz','SachsenAnhalt','Ahlum', 'Sangerhausen', 'Ramsin', 'Reinsdorf','Bernburg', 'Bitterfeld', 'Blankenburg', 'Born', 'Delitz', 'Derben', 'Dessau', 'Freist', 'Glinde', 'Halberstadt', 'Halle', 'Koethen', 'Loburg', 'Magdeburg', 'Mannhausen', 'Naumburg', 'Oschersleben', 'Roitzsch', 'Wernigerode'
]}

\clearpage
\section{Results of Regional Analysis}

\begin{table}[!ht]

\begin{center}
\begin{tabular}{|l c c c|} 
 \hline
 Province & Schwurbelarchiv & Mohr & Zehring \\ [0.5ex] 
 \hline\hline
Vienna & 11 & 33 & 1 \\ 
 \hline
 Styria &4 & 7& 0 \\
 \hline
 Vorarlberg & 6 & 10 & 0 \\
 \hline
 Salzburg & 4 & 7 & 0\\
 \hline
 Tyrol & 13 & 17 & 0 \\ 
  \hline
 Burgenland & 5 & 4 & 0 \\
  \hline
Carinthia & 0 & 7 & 0 \\
  \hline
 Upper Austria & 3 & 19 & 0 \\
  \hline
 Lower Austria & 1 & 5 & 0\\
 \hline
Austria & 14 & 88 & 1 \\
\hline\hline
\textbf{Total}&  61&197 & 2 \\[1ex] 
 \hline
\end{tabular}
\caption{Number of Telegram chats per dataset, related to each Austrian province, and Austria as a whole.}
\label{tab:province_counts_austria}
\end{center}

\end{table}

\begin{table}[!ht]

\begin{center}
\resizebox{\columnwidth}{!}{%
\begin{tabular}{|l c c c|} 
 \hline
 Province & Schwurbelarchiv & Mohr & Zehring \\ [0.5ex] 
 \hline\hline
Baden-Württemberg & 167 & 170 &  2\\ 
 \hline
Bavaria &109 & 147& 1 \\
 \hline
Berlin & 100& 136 &2 \\
 \hline
Brandenburg &133  & 87&5\\
 \hline
Bremen &  29& 19& 0 \\ 
  \hline
Hamburg &35 & 34 & 0 \\
  \hline
Hessen & 97 & 126& 1 \\
  \hline
Mecklenburg-Vorpommern & 59 & 65& 1\\
  \hline
Niedersachsen & 81& 82& 3\\
 \hline
Nordrhein-Westfalen & 174& 222& 1 \\ 
  \hline
Rheinland-Pfalz & 79& 103 & 0\\
  \hline
Saarland &30& 18& 1 \\
  \hline
Sachsen & 108 &136 &2 \\
  \hline
Sachsen-Anhalt &46 & 82& 1\\
 \hline
Schleswig-Holstein &51 & 55& 0\\
 \hline
Thüringen&45 &72 & 1 \\
 \hline
Germany&  171& 405 & 9 \\
 \hline\hline
\textbf{Total}&  1514& 1959 & 30 \\[1ex] 
 \hline
\end{tabular}
}
\caption{Number of Telegram chats per dataset, related to each German state, and Germany as a whole. }
\label{tab:province_counts}
\end{center}

\end{table}

\begin{table}[!ht]
\centering
\resizebox{\columnwidth}{!}{%
\begin{tabular}{|l c c c|}
\hline
City       & Schwurbelarchiv & Mohr & Zehring \\  [0.5ex] \hline\hline
Zürich               &             2            &      0               &           0       \\ \hline
Genf                 &               0          &     0                &          0        \\ \hline
Basel                &             0            &      5               &         0         \\ \hline
Lausanne             &           0              &        0             &         0         \\ \hline
Bern                 &          1               &         9            &         0         \\ \hline
Winterthur           &           0              &            1         &         0         \\ \hline
Luzern               &        1                 &            2         &         0         \\ \hline
St. Gallen           &        1                 &            0         &          0        \\ \hline
Lugano               &           0              &        0             &          0        \\ \hline
Biel/Bienne          &             1            &            0         &          0        \\ \hline
Switzerland        &             69          &           102        &          1      \\ \hline\hline
\textbf{Total}        &             75          &           119        &          1      \\ [1ex] \hline
\end{tabular}
}
\caption{Number of Telegram chats in Switzerland's 10 largest cities across different datasets. }
\label{tab:switzerland}
\end{table}

\end{document}